\documentclass[a4paper,11pt]{article}
\usepackage{pos}
\usepackage{booktabs}
\usepackage{multicol}
\usepackage{xspace}
\usepackage{ifthen}
\usepackage{heppennames2}
\usepackage{ptdr-definitions}
\usepackage{lipsum}

\title{CMS Upgrades for the High-Luminosity LHC Era}
\ShortTitle{CMS Upgrades for the High-Luminosity LHC Era}

\author*{Thiago Rafael Fernandez Perez Tomei}%
\onbehalf{on behalf of the CMS Collaboration}

\affiliation{Universidade Estadual Paulista,\\
  São Paulo, SP, Brazil}

\emailAdd{Thiago.Tomei@cern.ch}

\abstract{The High-Luminosity LHC (HL-LHC) era, set to begin in 2029,
will provide the general-purpose experiments with 
an instantaneous luminosity of up to \mbox{$\mathcal{L} = 7.5 \times 10^{34}$ cm$^{-2}$ s$^{-1}$} from pp collisions
at a centre-of-mass energy of 14\TeV.
To fully exploit this unprecedented data set,
the experimental setups must be upgraded to withstand the challenging conditions of the HL-LHC,
including up to 200 simultaneous collisions per bunch crossing and
a substantial radiation dose delivered to the detectors.
The CMS collaboration is currently undertaking the Phase-2 upgrade, 
which aims to enhance the detector's capabilities to maintain high performance under these conditions.
This upgrade includes
significant improvements to the muon spectrometer and barrel calorimeter,
a complete replacement of the silicon tracker, endcap calorimeter and beam radiation and luminosity subsystems,
the introduction of a new MIP timing detector layer,
and a redesigned trigger and data acquisition system.
These enhancements will ensure that the CMS experiment can fully profit from on the HL-LHC data, 
maximising its physics potential and 
expanding its ability to make the most precise measurements.}

\FullConference{12th Large Hadron Collider Physics Conference  (LHCP2024)\\
3-7 June 2024 \\
Boston, USA \\}

\newcommand{\pileupof}[1]{$\langle{\rm PU}\rangle = #1$\xspace}

\renewcommand{\logo}{\relax}

\begin{document}
\maketitle

\section{Introduction and the High-Luminosity LHC}

The High-Luminosity LHC (HL-LHC)~\cite{ZurbanoFernandez:2020cco} represents the next stage in the field of high-energy physics.
Reaching a centre-of-mass energy of 14\TeV and
achieving an instantaneous luminosity that can reach up to $\mathcal{L}=\text{7.5}\times\text{10}^\text{34}\,\text{cm}^\text{$-$2}\,\text{s}^\text{$-$1}$\xspace,
unprecedented for a hadron collider,
the HL-LHC will play an important role in the precision measurements of the standard model (SM) parameters.
It will also be instrumental not only in searches for rare SM processes,
such as $\PH \to \PGm\PGm$ and double Higgs boson production,
but also in searches for physics beyond the standard model.
The HL-LHC era will start with Long Shutdown 3 (2026--2028), 
a three-year period where the LHC accelerator chain will undergo improvements in order to reach the high-luminosity configuration;
physics operation is foreseen to extend up to 2041, delivering an integrated luminosity of up to $4000\fbinv$.

\section{The CMS Phase-2 Upgrade}

In order to adapt to the HL-LHC conditions, the CMS experiment~\cite{CMS:2008xjf} will undergo the so-called
\emph{Phase-2 upgrade}~\cite{CMSCollaboration:2015zni}.
The main goal is to extend the physics programme to the full HL-LHC data set whilst
keeping the detector performance in terms of selection efficiency, signal resolution and background rejection.
In order to achieve that goal, there are a number of challenges that have to be addressed.
The high instantaneous luminosity translates to a large number of quasi-simultaneous pp interactions, known as pileup (PU).
For the ultimate HL-LHC configuration, an average pileup \pileupof{200} is expected;
the upgraded detectors will have improved granularity and timing capabilities to mitigate this.
Conversely, the high integrated luminosity implies a large radiation dose delivered to the detectors;
a new silicon tracker and endcap calorimeter systems will be installed to be able to cope with those conditions.
Finally, a large fraction of the on-detector electronics,
as well as entire trigger and data acquisition system,
will undergo a complete overhaul to address the larger event size and acquisition rate during the HL-LHC era.

\subsection{Upgrade of the Tracker System}
\label{sec:tracker}

\begin{figure}[htbp]
   	\centering
	\includegraphics[height=20ex]{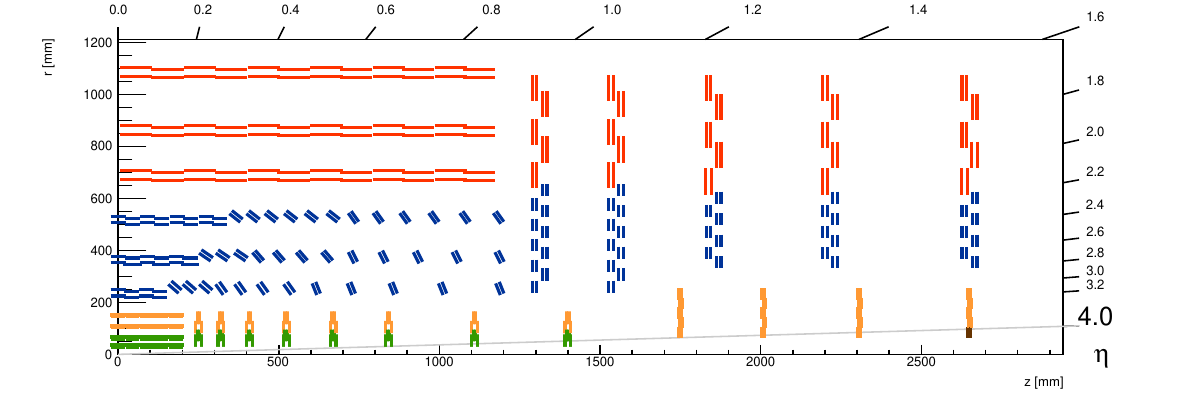}
	\raisebox{5ex}{\includegraphics[width=0.30\textwidth]{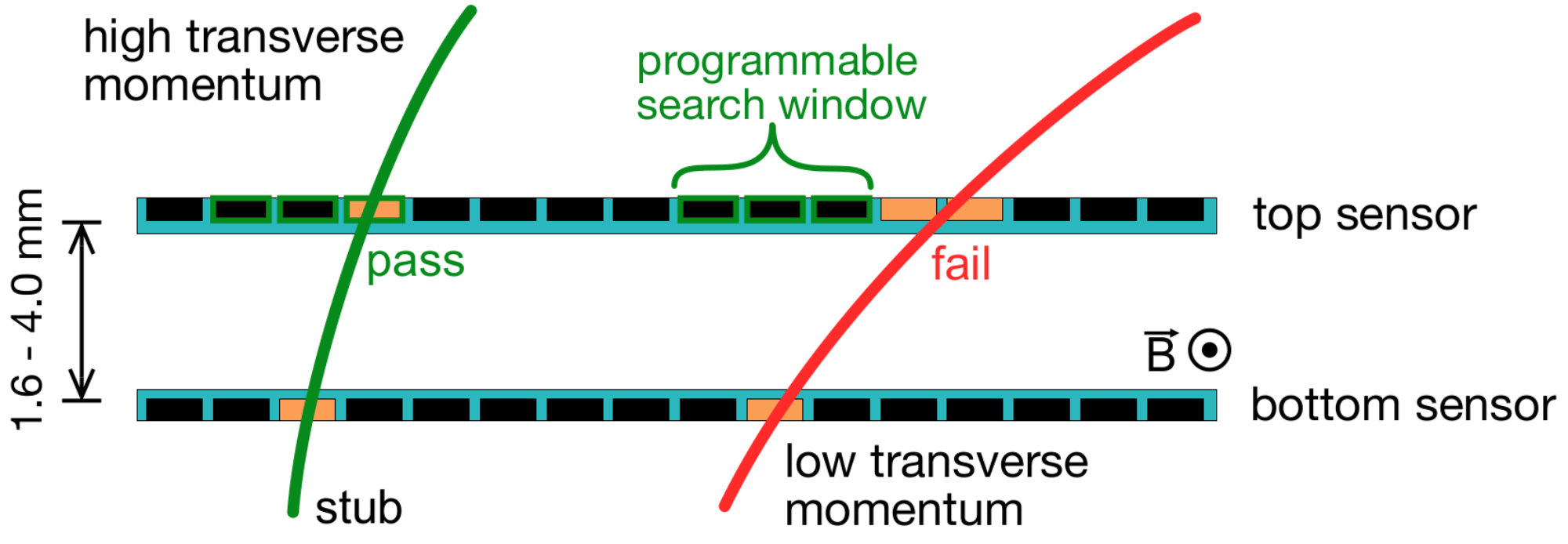}}
   	\caption{Left: CMS Phase-2 Tracker System geometry, divided in 
	Inner Tracker (orange and green modules)
	and
	Outer Tracker (red and blue modules).
	Right: Reconstruction of \PT modules in the Outer Tracker, which are input to the Level-1 Trigger decision.}
   	\label{fig:trackerGeometryAndPtModules}
\end{figure}

The Phase-2 Tracker System will again be divided in \emph{Inner Tracker} and \emph{Outer Tracker} systems,
featuring a symmetrical barrel + two endcaps geometry with acceptance extended up to $|\eta| < 4.0$~\cite{CMS:2017lum}.
The Inner Tracker will be pixel-based, with a pixel size of $25\times100\micron^2$,
organised in
four barrel layers and
eight small and four large discs per side,
totalling $2\times10^9$ channels.
The Outer Tracker will feature
44 million strips and 174 million macropixels,
organised in
six barrel layers and
five discs per side.
The design of the Tracker addresses the need for
radiation resistance and increased granularity,
allowing for the reconstruction of $\sim$1200 tracks per unit of $\eta$.
It also has a much reduced material budget in front of the calorimeters, preserving their resolution.
Finally, in contrast to its Phase-1 predecessor, the Phase-2 tracker will contribute to the Level-1 Trigger (L1T) decision through the Outer Tracker \PT modules, which are track stubs compatible with trajectories of particles with $\PT > 2\GeV$,
as discussed in Section~\ref{sec:tridas}.
Figure~\ref{fig:trackerGeometryAndPtModules} shows the Tracker geometry and the logic for the reconstruction of the \PT modules.

\subsection{The High-Granularity Calorimeter}

\begin{figure}[htbp]
   	\centering
	{\includegraphics[height=20ex]{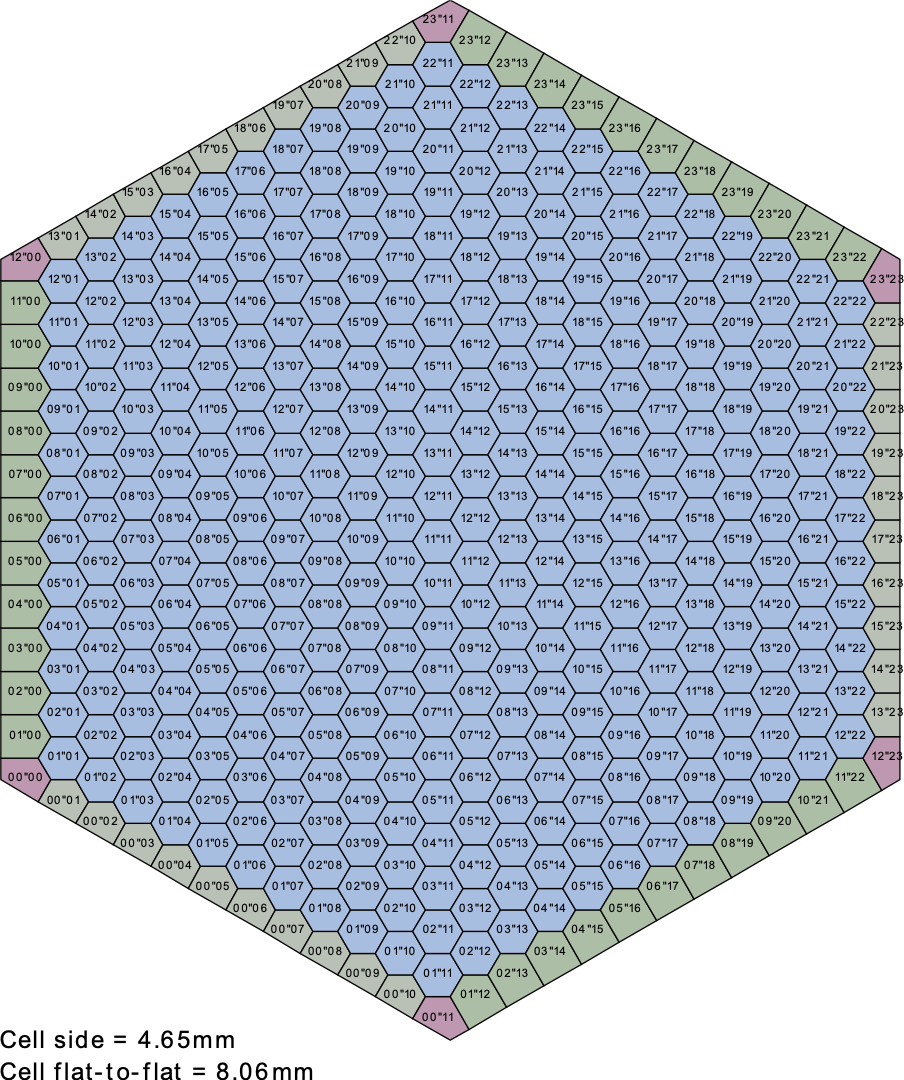}}\hspace{2em}
	{\includegraphics[height=20ex]{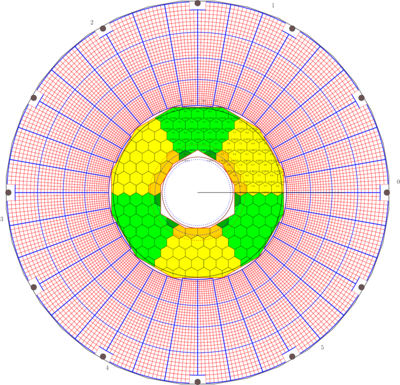}}\hspace{2em}
	{\includegraphics[height=20ex]{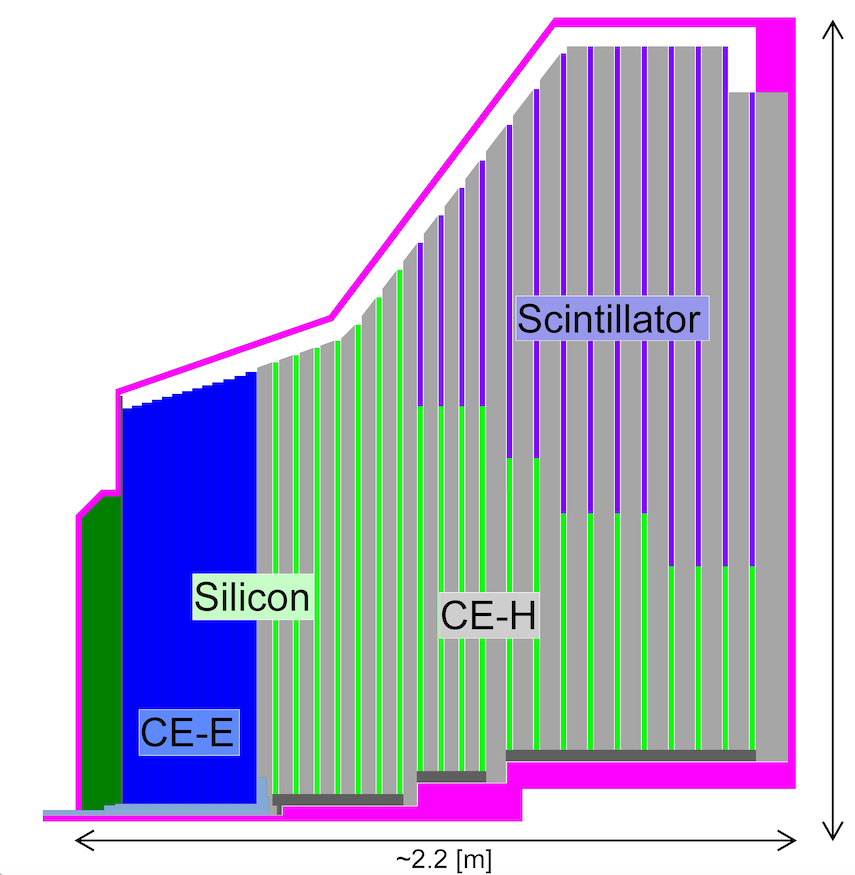}}
   	\caption{Left: arrangement of silicon sensors into an hexagonal module.
			 Middle: layout of layer 40, with hexagonal modules and scintillator cells.
             Right: HGCAL geometry overview, showing layers that comprise the CE-E and CE-H sections in an $r-z$ view.}
   	\label{fig:HGCALGeometry}
\end{figure}

The new endcap calorimetry system for the CMS Phase-2 upgrade has to address a number of requirements:
it has to be a dense calorimeter, to allow for high compactness of the particle shower's lateral development,
whilst displaying fine lateral and longitudinal granularity;
the capability of precisely measuring the time of the showers is also required.
Finally, the endcap calorimeter has to contribute to the L1T decision.
The \emph{High-Granularity Calorimeter} (HGCAL)~\cite{collaboration:2017gbu} is the proposed design to address those requirements.
It is a 5D calorimeter (position, energy and time) with coverage in the $1.5 < |\eta| < 3.0$ region,
segmented into 47 layers, and divided into
electromagnetic (CE-E) and hadronic (CE-H) sections,
with both sections structured as sampling calorimeters.
It uses two types of sensors:
silicon cells in CE-E and the high-radiation part of CE-H, and
scintillator cells elsewhere.
The silicon cells are grouped into hexagonal modules that are in turn grouped into ``cassetes'',
which provide cooling infrastructure, electronics and absorber elements.
The cassetes and the scintillator cells are finally grouped into the layers that compose the HGCAL.
Figure~\ref{fig:HGCALGeometry} shows the HGCAL geometry overview and sensors' arrangement.	

\begin{figure}[htb]
	\centering
	{\includegraphics[height=20ex]{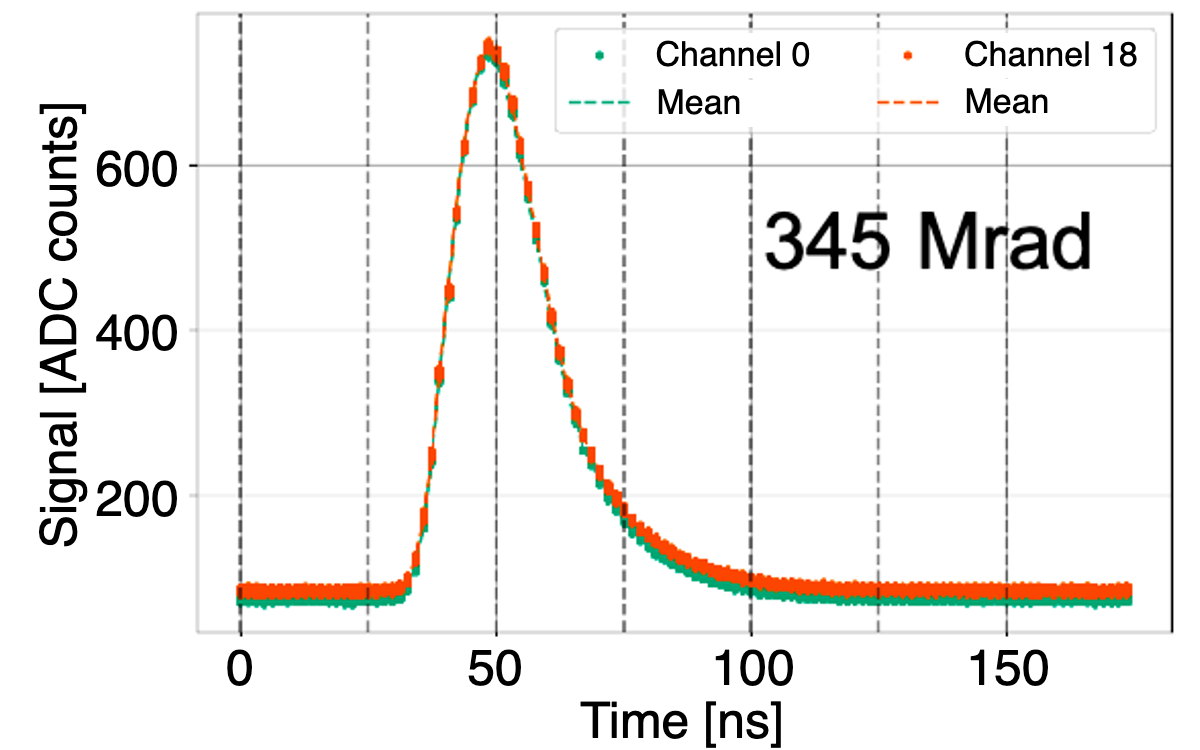}}\hspace{2em}
	{\includegraphics[height=20ex]{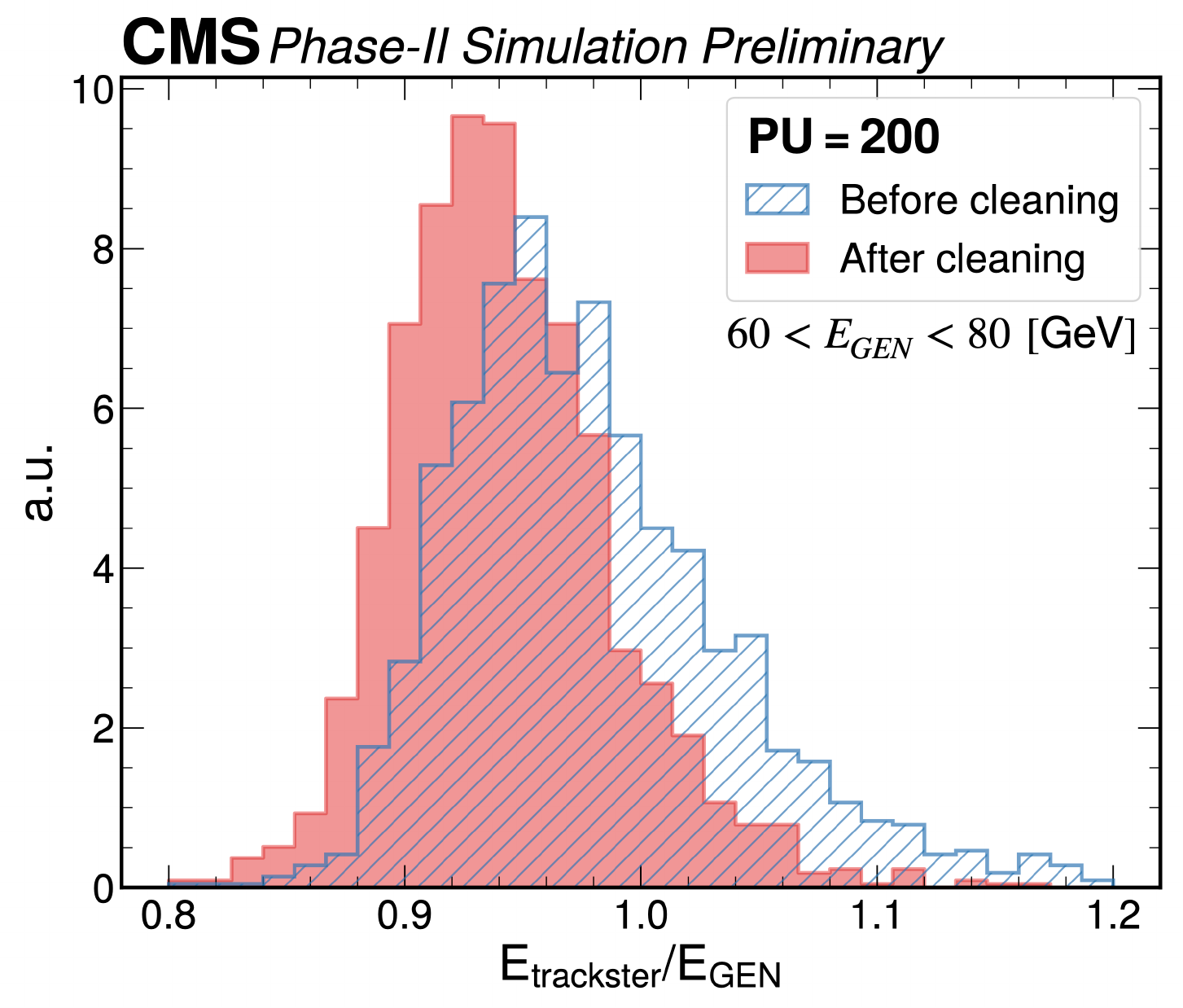}}
   	\caption{Left: The HGCROC ADC signal peak is well-behaved and stable even after 3.45\,MGy irradiation.
			Right: As an example of the HGCAL performance, it displays good reconstruction quality of 60--80\GeV photons, even without dedicated energy corrections but after a PCA-based cleaning of the shower constituents. 
}
   	\label{fig:HGCROCAndPerformance}
\end{figure}

The CE-E section comprises
26 layers of silicon sensors interspersed with Cu, CuW and Pb absorbers,
totalling 27.7\,$\chi_0$ and $\sim$1.5\,$\lambda$.
On the other hand, the CE-H section comprises
7 layers of Si sensors and 14 layers of mixed silicon--scintillator sensors,
interspersed with stainless steel and Cu absorbers and
totalling $\sim$8.5\,$\lambda$.
In total the HGCAL contains
approximately 6 million silicon and 240,000 scintillator channels, with each silicon (scintillator) sensor having $0.6$ or $1.2\cm^2$ (4--30$\cm^2$) cell size.
The HGCAL ReadOut Chip (HGCROC) is responsible for sensor readout,
providing fast signal shaping (signal peak in $<$ 25\,ns) and
$\sim$25\,ps timing resolution, whilst being radiation tolerant up to 2\,MGy.
The left panel of Fig.~\ref{fig:HGCROCAndPerformance} shows 
the HGCROC signal shape;
the right panel shows, as an example, the HGCAL performance on the reconstruction of 60--80\GeV photons.

\subsection{The MIP Timing Detector}

\begin{figure}[htbp]
	\centering
	{\includegraphics[height=20ex]{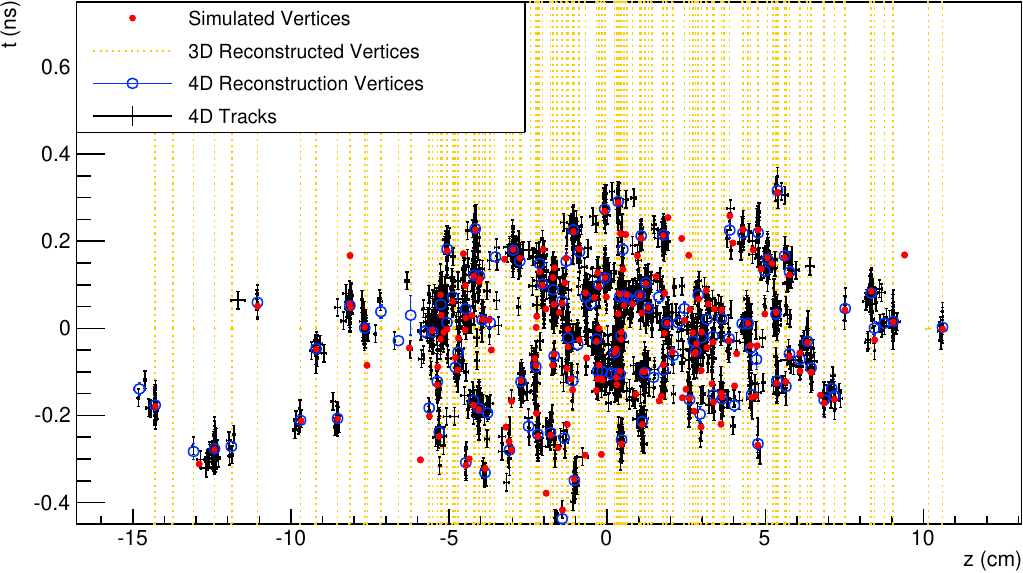}}
	{\includegraphics[height=20ex]{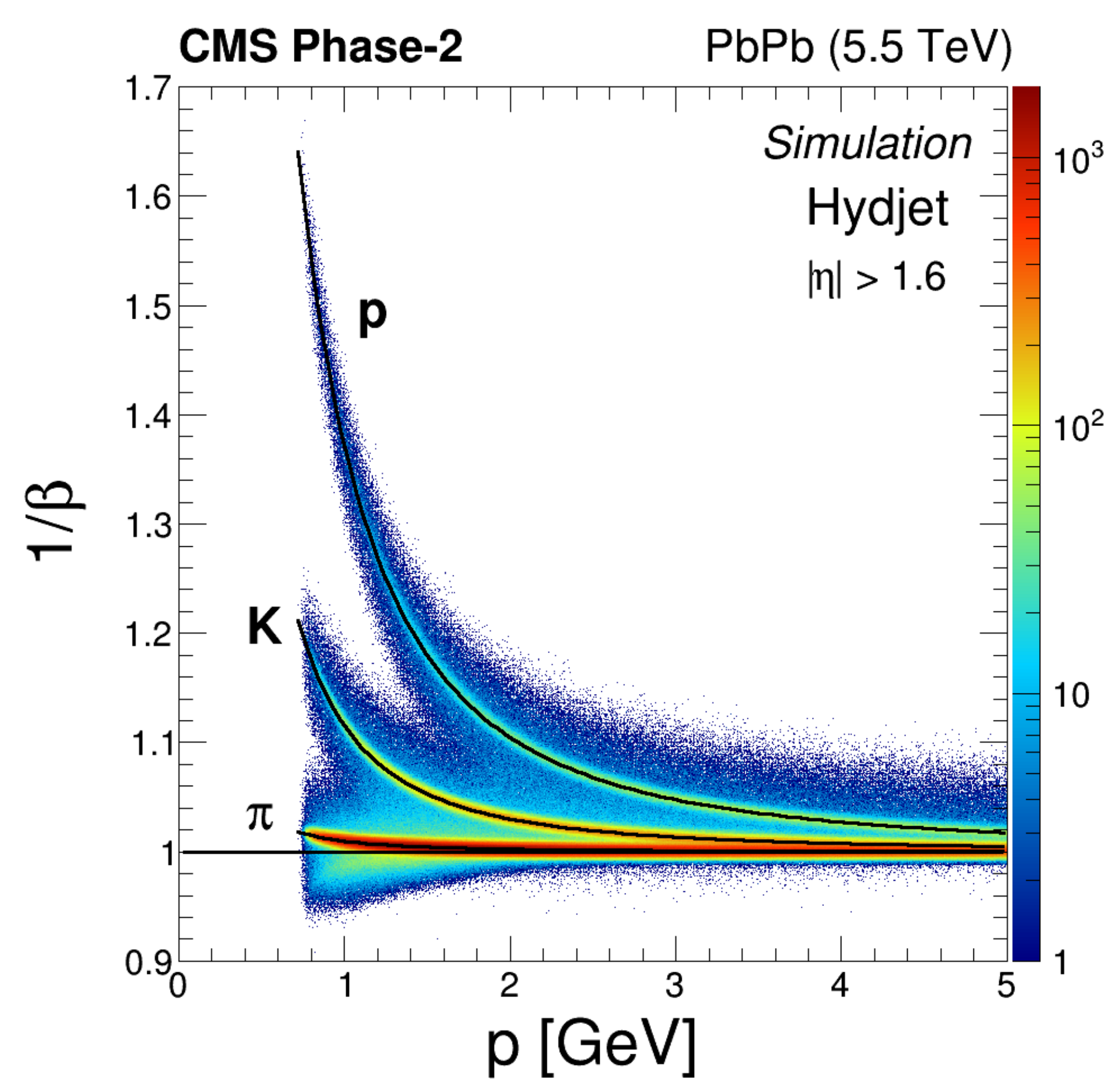}}
   	\caption{Left: the pp interactions are spread both in the longitudinal coordinate and in time.
	Leveraging the separation in the two dimensions leads to a better identification of the pileup vertices.
			 Right: particle identification of protons, kaons and pions with the ETL.}
   	\label{fig:MTDverticesAndPID}
\end{figure}

The \emph{MIP Timing Detector} (MTD) is a specialised tracking layer that aims to
measure the production time $t_0$ of minimum ionising particles~\cite{Butler:2019rpu}.
One of its main motivations is pileup mitigation:
since LHC proton bunches are spread longitudinally,
the pp interactions in a given bunch crossing are spread not only in that dimension but also in time,
with a root mean square close to 200\,ps.
So the identification of pileup interactions can be done in 2D, as shown in the left panel of Fig.~\ref{fig:MTDverticesAndPID}.
The MTD will also play a role in new physics searches, for instance increasing the identification efficiency of delayed particles, the time-of-flight measurement of heavy stable charged particles, amongst other possibilities.
Finally, the MTD will be a great asset for the heavy ions running in the HL-LHC era,
thanks to the particle identification capability it brings to the table, as shown in the right panel of Fig.~\ref{fig:MTDverticesAndPID}.

\begin{figure}[htb]
	\centering
	{\includegraphics[height=20ex]{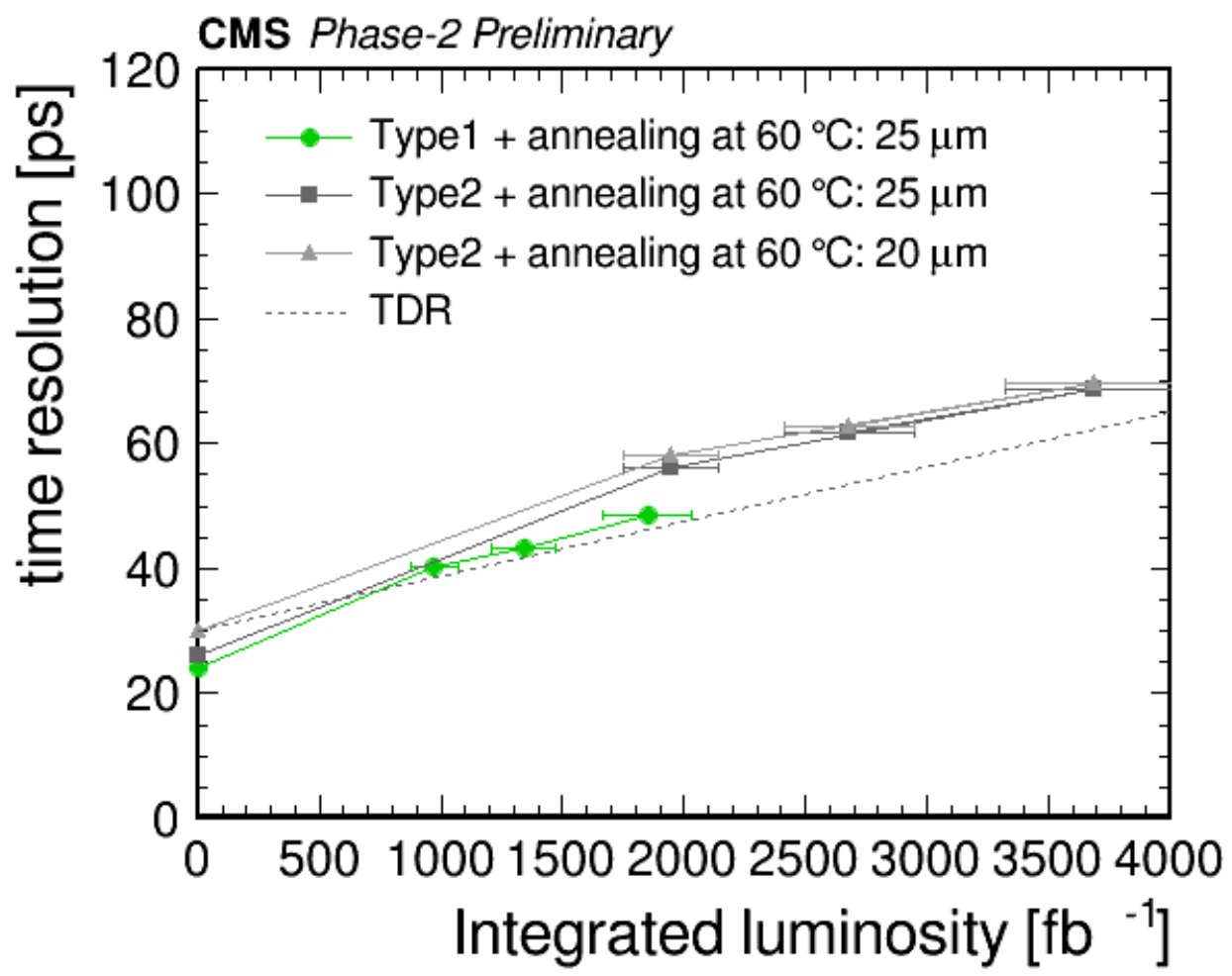}}
	{\includegraphics[height=20ex]{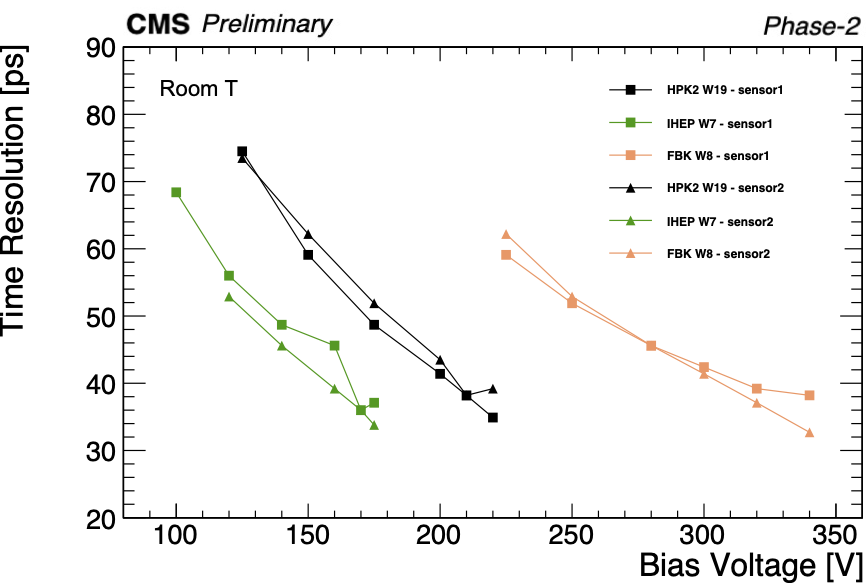}}
   	\caption{Left: BTL time resolution as a function of the equivalent integrated luminosity, for different types of modules. 
	The final choice for BTL is to use modules with 3.75\mm crystal width (``Type1'') and 25\micron cell-size SiPMs.
	Right: time resolution of $16\times16$ LGADs for ETL, measured with a 4--5\GeV electron test beam,
	showing time resolution lower than 40\,ps for bias voltages $<$ 350\,V.}
   	\label{fig:MTDResults}
\end{figure}

The MTD is divided in two sections: the
Barrel Timing Layer (BTL),
a cylindrical detector located inside the Tracker support tube, 
covers $|\eta| < 1.45$; the
Endcap Timing Layer (ETL),
a pair of disks located in front of the HGCAL thermal screen,
 covers the $1.6 < |\eta| < 3.0$ region.
The BTL sensors are LYSO:Ce scintillating crystal bars, 
instrumented with 25\micron cell-size silicon photomultipliers (SiPMs) on both ends of the crystal for readout.
16 crystals form a BTL array, and two arrays form a BTL module, encased by a copper housing. 
It contains $\sim$332k channels and has a total surface area of approximately 38\,$\text{m}^2$.
On the other hand, the ETL sensors are silicon low-gain avalanche diodes (LGADs),
grouped in $16\times16$ arrays and
bump-bonded to a custom readout chip.
The ETL contains $\sim$8.5M channels and has a total surface area of approximately 14\,$\text{m}^2$.
The timing resolution for the MTD reaches 30--65\,ps in the barrel region, and
35\,ps per track in the endcap.
Figure~\ref{fig:MTDResults} shows some of the latest characterisation studies for the MTD components.

\subsection{Upgrade of the Barrel Calorimeters and Muon Detectors}

\begin{figure}[htb]
	\centering
	{\includegraphics[height=20ex]{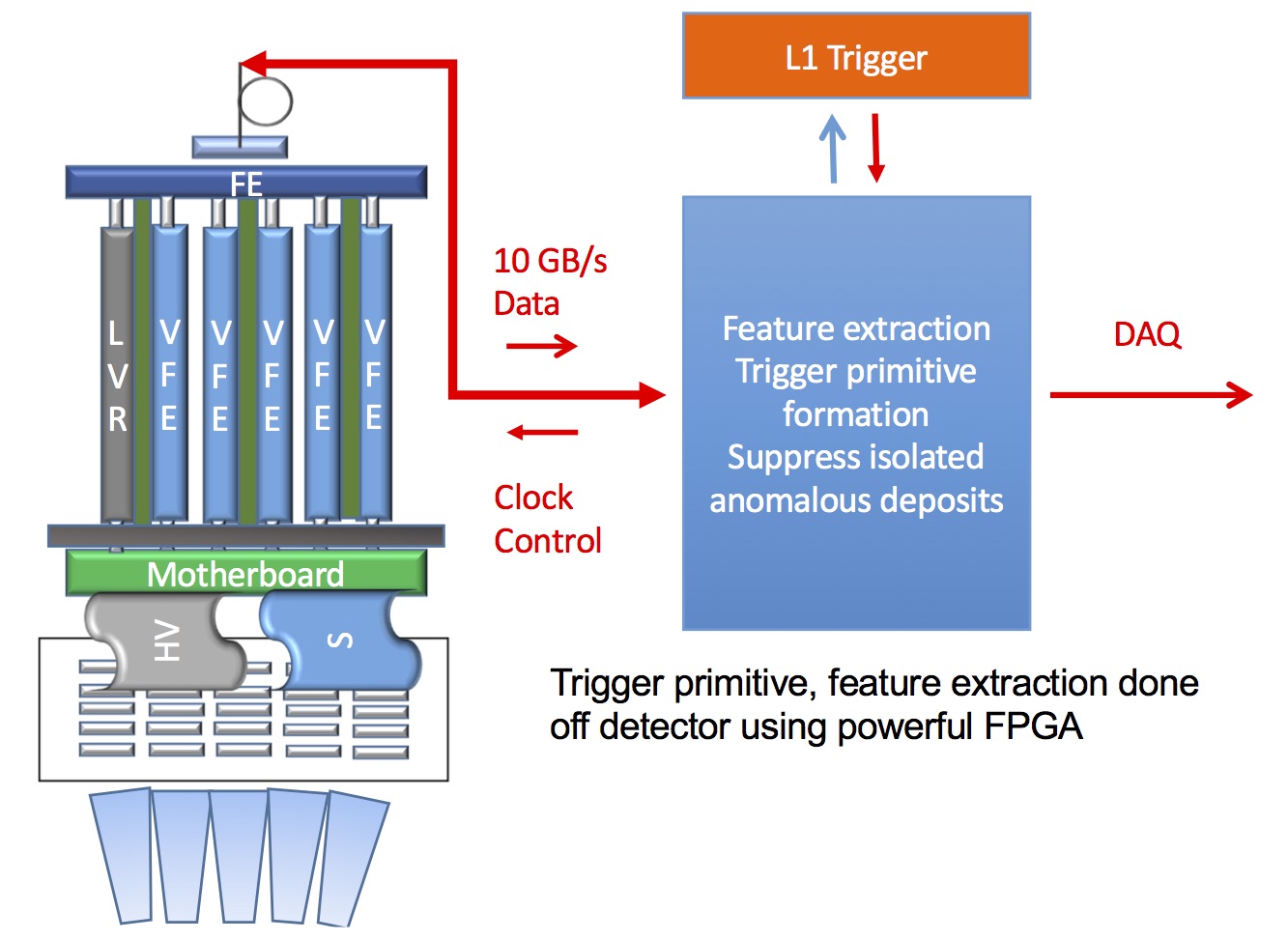}}\hspace{4em}
	{\includegraphics[height=20ex]{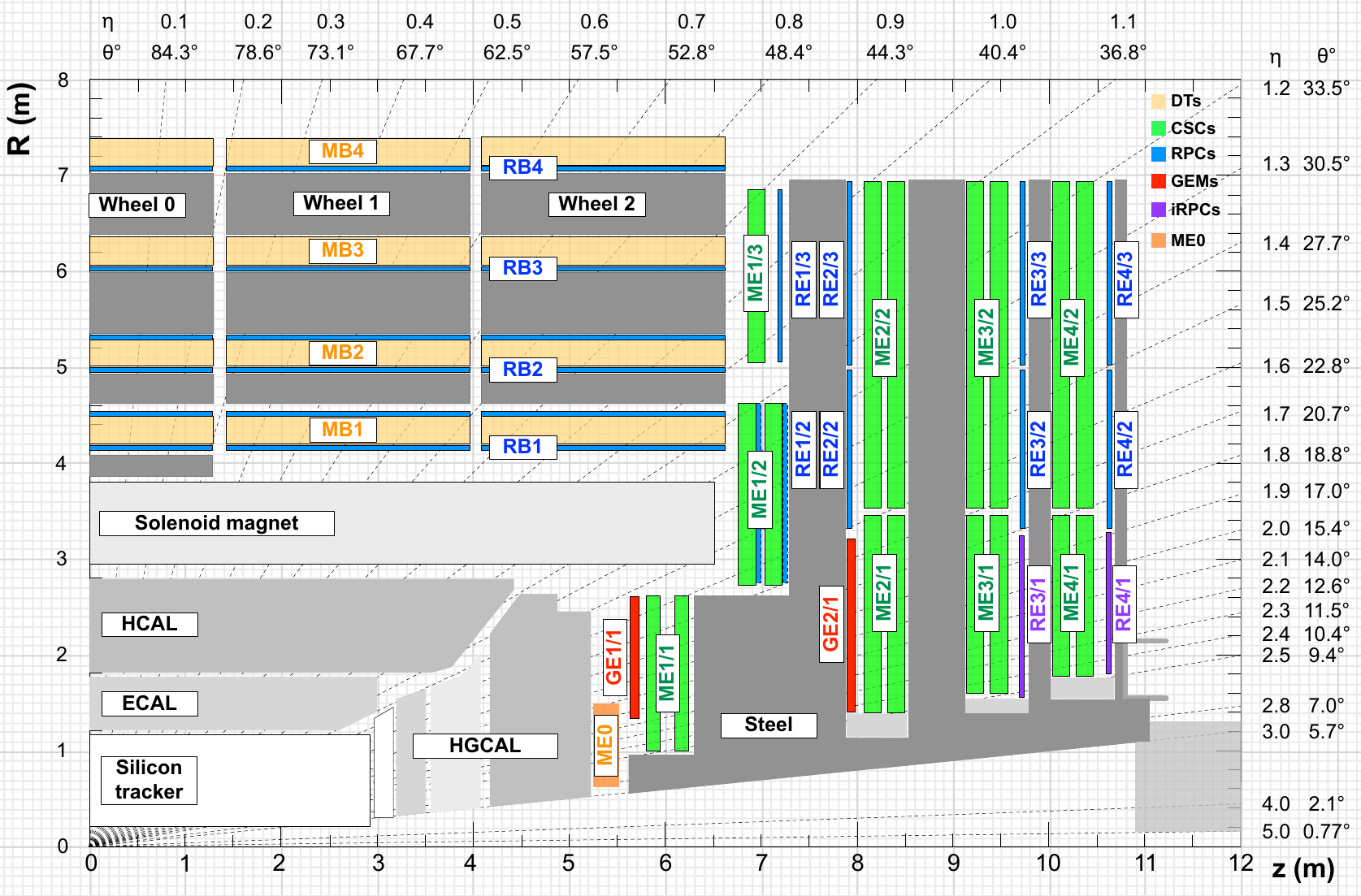}}
   	\caption{Left: Block diagram describing the upgraded EB electronics architecture.
	Right: overview of the CMS Phase-2 muon systems, with the new GEM chambers (ME0, GE1/1 and GE2/1) and iRPCs (RE3/1 and RE4/1).
	The new stations will enhance CMS's detection capabilities of forward muons.}
   	\label{fig:BCALandMuon}
\end{figure}

Besides the all-new endcap calorimeter and MIP timing detector,
the other CMS subsystems will also undergo improvements for the CMS Phase-2.
In order to address the extended latency and higher trigger rate requirements of the HL-LHC era (cf Section~\ref{sec:tridas}),
the CMS barrel calorimeters will undergo upgrades in their electronics systems~\cite{CERN-LHCC-2017-011}.
The electromagnetic barrel calorimeter (EB) and its hadronic counterpart (HB) will both deploy
a new back-end based around the \emph{Barrel Calorimeter Processor} (BCP) board.
The BCP board will be compliant with the Advanced Telecom Computing Architecture (ATCA),
including CMS-specific modifications,
and will be common to both sections, with the differences handled by firmware.
Each BCP will feature ``Linux-on-board'' for configuration,
 96 optical data links at 10.24 Gbps to EB 
(68 optical data links at 5.0   Gbps to HB),
and
distribution of the system clock from the LHC to the detectors 
with a time-interval error of the jitter standard deviation smaller than 10\,ps.
To fully support the EB system a total of 108 BCPs will be needed;
the HB system will deploy 18 BCPs, and additional specialised sections of the hadronic calorimeter
(Hadron Outer and Hadron Forward) will deploy 9 and 10 BCPs respectively.
Finally, the EB section will also feature upgrades to the front-end electronics,
with key points being 
the reduction of the signal shaping time and
the capability to deliver single-crystal information to the L1T.
The left panel of Fig.~\ref{fig:BCALandMuon} shows a block diagram
of the upgraded EB electronics architecture.

The muon spectrometers will also undergo Phase-2 upgrades~\cite{Hebbeker:2017bix}. 
The existing 
drift tubes (DT), 
cathode strip chambers (CSC) and 
resistive plate chambers (RPC) 
will all undergo upgrades of their electronic components to address the HL-LHC conditions;
a number of new muon chambers are also being built for the upgrade, in order to enhance the detection of forward muons.
Those include 
two \emph{improved RPCs} (iRPC) stations, 
planned to provide coverage in the $1.9 < |\eta| < 2.4$ range.
Those new chambers will have lower electrode thickness, smaller gas gaps, lower charge thresholds and integrated readout strips, with 
those enhancements allowing the iRPCs to deliver
better intrinsic timing resolution of 0.5\,ns and
better space resolution of 1.5\cm in $\eta$ (0.3--0.6\cm in $\phi$).
Also foreseen are three muon new stations based on the \emph{gas electron multiplier} (GEM) technology,
with an \mbox{Ar/CO$_2$} gas mixture and a triple GEM foil for the sensitive elements.
The ME0 GEM station will provide coverage in the $2.0 < |\eta| < 2.8$ range;
another GEM station, GE1/1, was installed during Long Shutdown 2 (2019--2021) and is already being used in data-taking.
The right panel of Fig.~\ref{fig:BCALandMuon} shows a geometry overview of the CMS Phase-2 muon systems.

\subsection{Upgrade of the Data Acquisition and Trigger Systems}
\label{sec:tridas}

The Data Acquisition (DAQ) is responsible for the data pathway from the detector readout, through the local storage at the experimental site, all the way to the transfer to offline storage.
It is also responsible for the time decoupling between the detector readout and the data reduction executed by the CMS Trigger Systems.
The experiment will continue using a two-level trigger system,
similarly to what {was} done for the first LHC era.
The Phase-2 Level-1 Trigger (L1T) will be implemented in custom-made electronics and dedicated to analyse the detector information
with a maximum latency of 12.5\,$\mu$s;
the Phase-2 High Level Trigger (HLT) will be implemented as a series of software algorithms, running in a heterogeneous-architecture computing farm.

\begin{figure}[htbp]
   \centering
   \includegraphics[height=24ex]{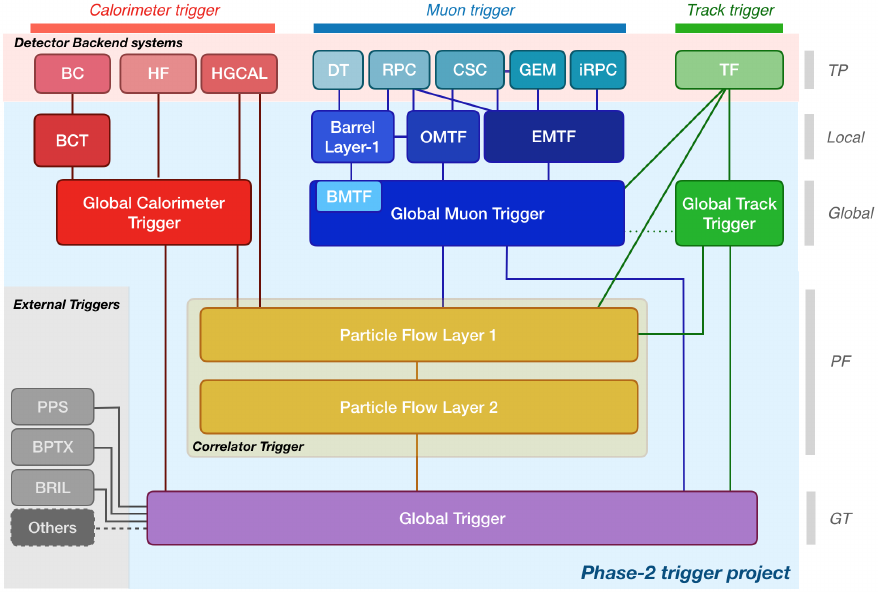}
   \includegraphics[height=24ex]{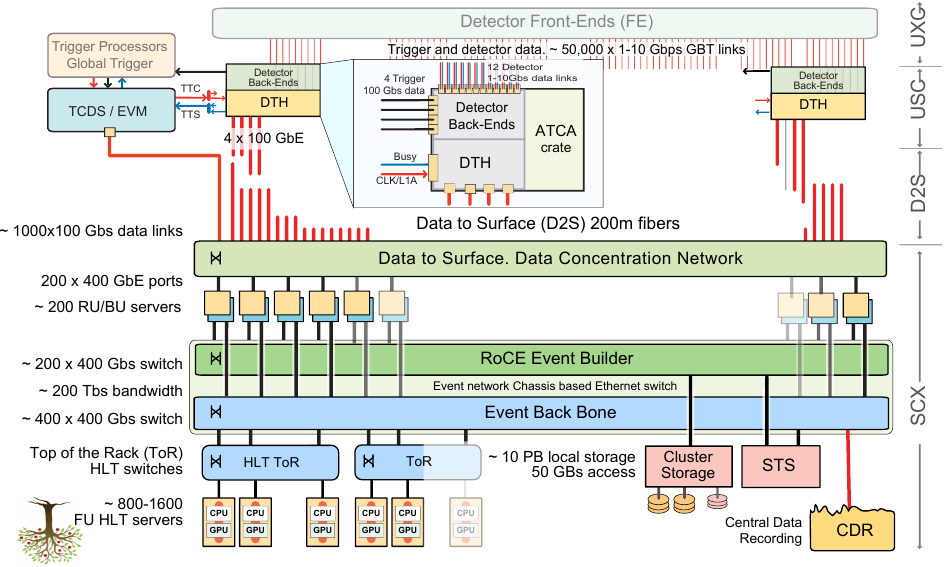}
   \caption{Left: Functional diagram of the CMS Phase-2 upgraded L1T design.
            Right: Conceptual structure of the CMS Phase-2 upgraded DAQ.}
   \label{fig:l1tdaq}
\end{figure}
The left panel of Fig.~\ref{fig:l1tdaq} shows the Phase-2 L1T architecture.
It will analyse the detector at a 40\,MHz rate, selecting events for further inspection at an output rate of up to 750\,kHz.
It will receive inputs from the calorimeters, the muon system and,
in the most significant change from Phase-1, from the Outer Tracker.
The \PT modules described in Section~\ref{sec:tracker} are used to seed Kalman-filter-based track reconstruction.
Extended tracking, without luminous region constraints, will also be done for displaced particles.
The L1T will also be equipped with a correlator layer that takes inputs from all subdetectors and runs advanced algorithms like particle flow reconstruction and machine learning particle identification.
Another highlight will be the presence of L1T scouting, making available to offline the intermediate L1T data streams, to be used for diagnostics, monitoring and physics.

The right panel of Fig.~\ref{fig:l1tdaq} shows the Phase-2 DAQ architecture.
The detector will be readout by $\sim$50,000 high-speed front-end optical links capable of sustaining up to 60 Tb/s data rate.
The experiment has standardised on the usage of the ATCA form-factor for the detector backends, and
a dual-function board DTH-400 is responsible both for DAQ data aggregation and timing and trigger control and distribution.
The event network will be implemented using remote direct memory access over Converged Ethernet (RoCE),
delivering completed events to the HLT at a rate of up to 750\,kHz,
with event sizes in the 7--10\,MB range.
The HLT system will be implemented on GPU-equipped servers,
with a combined computing power of 37\,MHS06 and an output rate and bandwidth of %
10\,kHz and 70--100\,GB/s respectively.
The DAQ will also contain a buffer of $\sim$10\,PB for local storage prior to final transfer to tape.
The complete information on the Phase-2 L1T and HLT is available the respective Technical Design Reports~\cite{Zabi:2020gjd,Collaboration:2759072}.

\subsection{The Phase-2 Beam Radiation, Instrumentation and Luminosity Systems}

The systems for monitoring beam radiation,
instrumentation
and luminosity (BRIL)
are all integrated within CMS.
The goal of the beam radiation and beam instrumentations sections is to
optimise the protection and lifetime of CMS's subdetectors
both in terms of real-time radiation exposition and integrated fluence.
Conversely, one of the main requirements of the luminosity monitoring system is
the capability for real-time measurements,
in order to allow optimisation of operational scenarios.
A second key parameter is the precision of the system,
which should reach 
$\sim$2\% precision during online operations
and
$\sim$1\% precision after offline calibrations.
Finally, the system should be able to deliver both bunch-by-bunch and fill-integrated luminosities.

To address these needs, the BRIL system will both upgrade its current components and deploy a number of new ones.
Amongst the latter we highlight a new dedicated data path coming from the Inner Tracker Endcap Pixel (TEPX) detector that 
will allow it be used for luminosity measurements, besides its role in track reconstruction. 
The innermost ring of the outermost disk of the TEPX, 
covering the region $|\eta| > 4.0$ in the right panel of Fig.~\ref{fig:trackerGeometryAndPtModules}
(coloured in brown),
will be exclusive to this data path.
Conversely, new systems based on the Outer Tracker \PT modules from the sixth barrel layer and on
the central muons reconstructed by the L1T 40\,MHz scouting system will also be deployed;
the latter is foreseen to provide the best statistical uncertainty in luminosity measurements during physics production.
A new detector, the \emph{Fast Beam Conditions Monitor},
based on silicon sensors and custom readout, 
is yet another addition to BRIL's collection of luminometers;
it will have the advantage of being available outside of stable beams,
while also being independent of both central trigger and DAQ systems.
Finally, on the beam radiation monitoring side, 
a pair of systems of neutron monitors are being developed, 
one based on gas-filled proportional counters and
the other on the combination of Bonner sphere neutron spectrometers with SiPM.
A full description is given in the BRIL system's Technical Design Report~\cite{Collaboration:2759074}.



\section{Conclusions}

The High-Luminosity LHC paves the way for the full exploitation of the accelerator,
completing the cycle that started with the Large Electron-Positron Collider.
The full luminosity of 4000\fbinv,
to be achieved by the end of the experimental run in the 2040s,
will be needed for the most extensive searches and the most precise SM measurements.
But that extensive data set will come at a price:
the HL-LHC experiments will have to address the harshest conditions in a hadron collider to date,
with up to 200 simultaneous pp collisions per bunch crossing,
subjecting the detectors to very high radiation doses and data acquisition rates.

The CMS Collaboration has prepared a roadmap to remodel their detector, 
the Phase-2 upgrade,
to be able to profit from the HL-LHC data set.
The detector will be equipped with three brand new subsystems:
a revamped silicon Tracker,
a five-dimensional Endcap Calorimeter,
and a MIP Timing Detector.
The other subsystems will also undergo extensive improvements, 
whilst keeping portions of the current hardware.
With the Phase-2 upgrade, CMS will be able to keep the detector's very high performance,
allowing the collaboration to fully achieve their physics programme
and
make the most of the data collected during the HL-LHC era.

\small

\providecommand{\href}[2]{#2}\begingroup\raggedright\endgroup


\begin{thebibliography}{10}

\bibitem{ZurbanoFernandez:2020cco}
I.~Zurbano~Fernandez et~al., \emph{{High-Luminosity Large Hadron Collider
  (HL-LHC): Technical design report}}, 
  \href{https://doi.org/10.23731/CYRM-2020-0010}{CERN Yellow Reports: Monographs, 10/2020}.

\bibitem{CMS:2008xjf}
{\scshape CMS} collaboration, \emph{{The CMS Experiment at the CERN LHC}},
  \href{https://doi.org/10.1088/1748-0221/3/08/S08004}{\emph{JINST} {\bfseries
  3} (2008) S08004}.

\bibitem{CMSCollaboration:2015zni}
{\scshape CMS} collaboration, ``{Technical Proposal for the Phase-II Upgrade of
  the CMS Detector}.'' \url{https://cds.cern.ch/record/2020886}, 2015.

\bibitem{CMS:2017lum}
{\scshape CMS} collaboration, ``{The Phase-2 Upgrade of the CMS Tracker}.''
  \url{https://cds.cern.ch/record/2272264}, 2017.

\bibitem{collaboration:2017gbu}
{\scshape CMS} collaboration, ``{The Phase-2 Upgrade of the CMS Endcap
  Calorimeter}.'' \url{http://cds.cern.ch/record/2293646}, 2017.

\bibitem{Butler:2019rpu}
{\scshape CMS} collaboration, ``{A MIP Timing Detector for the CMS Phase-2
  Upgrade}.'' \url{http://cds.cern.ch/record/2667167}, 2019.

\bibitem{CERN-LHCC-2017-011}
{\scshape CMS} collaboration, ``{The Phase-2 Upgrade of the CMS Barrel
  Calorimeters}.'' \url{https://cds.cern.ch/record/2283187}, 2017.

\bibitem{Hebbeker:2017bix}
{\scshape CMS} collaboration, ``{The Phase-2 Upgrade of the CMS Muon
  Detectors}.'' \url{https://cds.cern.ch/record/2283189/}, 2017.

\bibitem{Zabi:2020gjd}
{\scshape CMS} collaboration, ``{The Phase-2 Upgrade of the CMS Level-1
  Trigger}.'' \url{https://cds.cern.ch/record/2714892}, 2020.

\bibitem{Collaboration:2759072}
{\scshape {CMS}} collaboration, ``{The Phase-2 Upgrade of the CMS Data
  Acquisition and High Level Trigger}.''
  \url{https://cds.cern.ch/record/2759072}, 2021.

\bibitem{Collaboration:2759074}
{\scshape CMS} collaboration, ``{The Phase-2 Upgrade of the CMS Beam Radiation
  Instrumentation and Luminosity Detectors}.''
  \url{https://cds.cern.ch/record/2759074}, 2021.

\end{thebibliography}
\end{document}